\documentclass{article}
\usepackage[T1]{fontenc}
\usepackage{official_template_2027/spconf,amsmath,amssymb,graphicx}
\usepackage{booktabs,multirow}
\usepackage{xcolor}
\usepackage[hidelinks]{hyperref}
\usepackage{balance,stfloats,needspace}
\newcommand{\weights}{\mathbf W}
\newcounter{paperalgorithm}

\title{AWAKEN, THEN SCALE: TINY ADAPTATION FOR MRI RECONSTRUCTION}
\name{Mohammed Wattad, Tamir Shor, and Alexander M. Bronstein}
\address{Faculty of Computer Science, Technion-Israel Institute of Technology, Haifa, Israel}
\begin{document}
\ninept
\raggedbottom
\maketitle
\begin{abstract}
Dormant Awakening (DA) specializes a magnetic resonance imaging (MRI)
reconstruction model by fitting a small set of zero weights to one labeled
slice. Across fourteen U-shaped convolutional network (U-Net) and vision
transformer (ViT) sources, fitting 1,004 or 10,560 weights gives mean
peak signal-to-noise ratio (PSNR) gains of .430 and .509 dB.
We analyze how scaling the fitted output correction changes evaluation
PSNR, then use the calibration correction size to cap changes on new
images. The cap bounds changes in root mean squared error (RMSE)
relative to the source without an evaluation reference.
At 10\% adaptation budget, we compare fixed halving and dynamic capping
for DA, unrestricted sparse adaptation and low-rank adaptation (LoRA).
The controls increase both measured calibration-evaluation correlations
across all six tested architecture-adapter settings. Experiments concern one constructed fastMRI-to-M4Raw shift with fixed
source checkpoints and participant panels.
\end{abstract}

\section{Introduction}

Learned MRI reconstruction can degrade when test acquisition or image
distributions differ from training~\cite{knoll2019,darestani2022ttt}.
Parameter-efficient adaptation addresses such shifts by updating only a
small part of a trained reconstructor~\cite{peaci}. Pruning sets selected network weights to zero~\cite{snip} and has also
been studied for MRI reconstruction~\cite{punit}, leaving dormant parameters that can
potentially be reused for adaptation.

A labeled MRI slice supplies an undersampled input and reference.
Dormant Awakening (DA) fits selected zero weights of a trained source
model to this calibration pair while keeping its nonzero weights fixed.
The fitted parameter update, called the \textbf{adapter}, stays fixed
during evaluation on other participants, and resetting it recovers the
source. We ask whether restricted updates improve reconstruction and
whether output control makes calibration loss informative about
evaluation.

Using pruned capacity for later adaptation is related to
PackNet~\cite{packnet}, PAC-Net~\cite{pacnet}, and pruning-based domain
adaptation~\cite{pacda}. FAM~\cite{fam,famjournal} adapts zeros after
sparse meta-training, including one-shot experiments, while TACT~\cite{tact}
adapts inactive or near-zero weights without labels. LoRA~\cite{lora},
parameter-efficient computational imaging~\cite{peaci}, and sparse
methods~\cite{pafi,sara,shira,saft,diffpruning,rosa} provide other compact
updates. Single-shot~\cite{arvinte2022} and self-supervised MRI
methods~\cite{ssdu,zsssl,darestani2022ttt,adaptNet,d2sa} address
reconstruction adaptation under different supervision and model settings.
Combining source and adapted models can also support robustness~\cite{wiseft}.

These works leave open whether dormant weights of an already trained
sparse MRI reconstructor provide a useful one-slice adaptation space
without adaptation-specific source training, and how the resulting
corrections should be controlled at inference. We contribute \textbf{(i) adaptation of few dormant weights}, compared
with unrestricted sparse adaptation and LoRA, \textbf{(ii) an exact
analysis of how correction alignment and scale determine MSE and PSNR},
including when a fitted correction overshoots, and \textbf{(iii) a
calibration-derived reference-free output cap} for all three adapters.

\section{Single-slice adaptation}
\label{sec:adaptation}

Dormant Awakening (DA) adapts a pruned reconstruction model by training a
small subset of its zero weights while keeping all originally nonzero weights
fixed. Let $\Omega_0$ denote the dormant pool of exact-zero coordinates in
the source weights $\weights_s$. DA selects a \textbf{support}
$D\subseteq\Omega_0$ of $Q=|D|$ trainable coordinates. The operator $P_D$
places update values $u$ at these coordinates and zeros elsewhere, so for
layer input $h$ the adapted layer computes $(\weights_s+P_Du)h$. This covers
linear layers and convolutions written as matrix operations. Initializing
$u=0$ preserves the source. A dense checkpoint stores the fitted values in
place, while a portable adapter stores their indices and values.

We use mean absolute error (MAE) as the calibration objective $\mathcal L$
and consider SNIP-based and random support selection.
\textbf{Single-shot network pruning} (SNIP)~\cite{snip} scores
$|w_i\partial\mathcal L/\partial w_i|$, which vanishes at zero.
Our probe-selected DA (DA-SNIP) inserts seeded Xavier-uniform values $p_i$
into a disposable copy and selects the top $Q$ scores
$|p_i\partial\mathcal L/\partial p_i|$, with active weights frozen. Ties
follow layer and row-major order. We discard the probes, then fit the
selected coordinates from zero, bypassing their pruning mask. Random DA
samples zeros uniformly without replacement.

\textbf{Unrestricted sparse adaptation} ($S$) selects the largest
calibration-loss gradient magnitudes at the source across eligible active and
dormant coordinates. Pruned zeros remain differentiable, with global ranking
and no layer quotas. At equal budget, the unrestricted class contains all DA updates, so its
exact calibration optimum is at least as good. Evaluation also depends
on optimization and transfer. In result tables, $\mathrm{D}$ denotes
DA-SNIP.

We also compare with LoRA, which adds a low-rank product of factors $A$
and $B$ to the frozen source weights. For each architecture, budgets are
0.1\%, 1\%, and 10\% of the mean eligible active-weight count of sources
obtained with PUN-IT~\cite{punit} and RigL~\cite{rigl}. PUN-IT learns a
pruning mask at initialization and then trains the sparse network, while
RigL dynamically updates its sparse topology during training. Matching
these budgets to feasible LoRA ranks gives $Q=1{,}004$, $10{,}039$,
$100{,}387$ for U-Net and $10{,}560$, $119{,}680$, $1{,}189{,}760$ for
ViT. Equal parameter counts can produce different output changes,
motivating output-scale control.

LoRA uses scaling equal to its rank. U-Net allocates integer ranks within
the parameter budget after lowering a shared minimum rank. ViT distributes
ranks in block and multilayer perceptron (MLP) map order, assigning
remainders first. Layer coverage is limited at the smallest budget.

For source calibration loss $L_0$ and adapted-output loss $L_a$, the
\textbf{normalized calibration gain} (NCG) is
$X=(L_0-L_a)/(L_0+L_a+10^{-12})$.
Both use calibration L1 loss. Larger $X$ indicates a better relative fit.
For a scaled or capped output, $L_a$ is the loss of that output rule. NCG is computed from the calibration pair and is the score whose
association with mean gain on held-out participants we test.

\section{Theoretical Analysis}
\label{sec:theory}
A correction fitted to one calibration slice may transfer with an
inappropriate magnitude to other images. We therefore separate fitting
from output scale: we first analyze how scaling a fixed correction changes
reconstruction error, then derive a reference-free limit on that correction.

\subsection{Output scaling}
For one image, let $f_0$ be the source reconstruction,
$f_{\rm adapted}$ the fitted reconstruction, and $y$ the reference.
Define the \textbf{residual} and adapter \textbf{correction}
\[
 r=y-f_0,\qquad d=f_{\rm adapted}-f_0.
\]
An \textbf{output scale} $a\ge0$ gives
\[
 f_a=f_0+ad,\qquad y-f_a=r-ad.
\]
Thus $a=0,.5,1$ give the source, half-scaled output (Half), and full
adapted output (Full), respectively. Our controls use $a\in[0,1]$.

Let $b=\|r\|_2^2/m$ be the source MSE and
$E=\|d\|_2^2/m$ the correction energy per pixel. Then
\begin{equation}
\begin{aligned}
 G(a)&=\frac{\|r\|_2^2-\|r-ad\|_2^2}{m}\\
     &=\frac{2a\langle r,d\rangle-a^2\|d\|_2^2}{m}.
\end{aligned}
\label{eq:transfer-mse-short}
\end{equation}
For $E>0$, define
\begin{equation}
 t=\frac{\langle r,d\rangle}{\|d\|_2^2},\qquad
 H(a)=a(2t-a),\qquad G(a)=EH(a),
 \label{eq:normalized-gain}
\end{equation}
where $t$ measures correction alignment relative to its energy.
Thus the sign of $H(a)$ determines whether MSE improves. For $E=0$,
we set $H(a)=0$.

Since $H'(a)=2(t-a)$, the optimum over our tested range is
\begin{equation}
 a^*=\min\{1,\max\{0,t\}\}.
\label{eq:optimal-scale}
\end{equation}
Without the upper bound $a\le1$, the optimum over $a\ge0$ is
$\max\{0,t\}$. For $t>0$, applying $a>t$ overshoots this optimum.
For $a>0$, $H(a)>0$ exactly when $a<2t$, so Half and Full improve
MSE for $t>1/4$ and $t>1/2$, respectively. These statements concern
MSE, while adapter fitting minimizes L1 loss. Computing $t$ requires
the reference $y$ and therefore cannot guide inference on a new patient.

For positive source and scaled-output MSE,
\begin{equation}
 \Delta\operatorname{PSNR}(a)
 =10\log_{10}\!\left(\frac{b}{b-EH(a)}\right).
 \label{eq:normalized-gain-psnr}
\end{equation}
For a fixed pair this increases with $H(a)$. Across pairs, $b$ and $E$
also vary, so we compute PSNR per image before averaging.

\subsection{Reference-free cap}
Without a reference image, the source and adapted outputs still determine
the correction size. For
$\operatorname{RMS}(v)=\|v\|_2/\sqrt m$, the reverse triangle inequality
gives
\begin{equation}
 |\operatorname{RMSE}(g,y)-\operatorname{RMSE}(f_0,y)|
 \le\operatorname{RMS}(g-f_0).
 \label{eq:rmse-cap-short}
\end{equation}
Thus limiting the RMS output change to $\tau\ge0$ bounds the corresponding
RMSE change. Let $R=\operatorname{RMS}(d)$. The \textbf{radial cap} is
\begin{equation}
 f_\tau=f_0+a_\tau d,\qquad a_\tau=\min\{1,\tau/R\}.
 \label{eq:output-cap-short}
\end{equation}
For $R=0$, we set $a_\tau=1$. The radial cap is the Euclidean projection
onto outputs within distance $\tau\sqrt m$ of $f_0$~\cite{boyd2004convex}.
Since mean absolute change is at most RMS, $\tau$ also bounds the absolute
change in MAE.
Writing $d=R\tilde d$ with $\operatorname{RMS}(\tilde d)=1$ gives
$t=A/R$, where $A=\langle r,\tilde d\rangle/m$ depends on the unknown
reference. Before clipping, the reference-informed optimal scale is $A/R$, while
Cap uses $\tau/R$. Thus both respond inversely to correction size, with
$\tau$ obtained from calibration rather than the unknown reference.
\subsection{Inference rules}
We apply two controls to the same fitted adapters: fixed half scaling
(Algorithm~\ref{alg:half}) and the image-dependent cap
(Algorithm~\ref{alg:cap}).
\par\medskip
\begingroup
\setlength{\fboxsep}{5pt}
\setlength{\fboxrule}{0.4pt}
\noindent\fcolorbox{black!35}{black!5}{%
\begin{minipage}{\dimexpr\columnwidth-2\fboxsep-2\fboxrule\relax}
\refstepcounter{paperalgorithm}\label{alg:half}
\textbf{Algorithm \thepaperalgorithm. Fixed-half inference}
\par\smallskip\hrule height 0.4pt\smallskip
\textbf{Input:} Frozen source $f_0$, fitted DA, $S$ or LoRA model
$f_{\rm adapted}$, and new image $x$.
\par\smallskip
\textbf{Output:} Reconstruction with half the fitted correction.
\begin{list}{\arabic{enumi}:}{%
\usecounter{enumi}
\setlength{\leftmargin}{1.25em}
\setlength{\labelwidth}{0.85em}
\setlength{\labelsep}{0.4em}
\setlength{\itemsep}{3pt}
\setlength{\parsep}{0pt}
\setlength{\topsep}{4pt}}
\item Compute $f_0(x)$ and $f_{\rm adapted}(x)$.
\item Return $f_0(x)+\tfrac12[f_{\rm adapted}(x)-f_0(x)]$.
\end{list}
\end{minipage}}
\endgroup
\par\medskip

\par\medskip
\begingroup
\setlength{\fboxsep}{5pt}
\setlength{\fboxrule}{0.4pt}
\noindent\fcolorbox{black!35}{black!5}{%
\begin{minipage}{\dimexpr\columnwidth-2\fboxsep-2\fboxrule\relax}
\refstepcounter{paperalgorithm}\label{alg:cap}
\textbf{Algorithm \thepaperalgorithm. Capped inference}
\par\smallskip\hrule height 0.4pt\smallskip
\textbf{Input:} Frozen source $f_0$, fitted DA, $S$ or LoRA model
$f_{\rm adapted}$, calibration input $x_c$, and fraction
$c\in[0,1]$ ($c=.5$ here).
\par\smallskip
\textbf{Output:} Capped reconstruction for each new image $x$.
\begin{list}{\arabic{enumi}:}{%
\usecounter{enumi}
\setlength{\leftmargin}{1.25em}
\setlength{\labelwidth}{0.85em}
\setlength{\labelsep}{0.4em}
\setlength{\itemsep}{3pt}
\setlength{\parsep}{0pt}
\setlength{\topsep}{4pt}}
\item Set the limit once. Measure
$R(x_c)=\operatorname{RMS}(f_{\rm adapted}(x_c)-f_0(x_c))$
and store $\tau=cR(x_c)$.
\item For each new image $x$, compute $f_0(x)$ and
$f_{\rm adapted}(x)$. Form
$d(x)=f_{\rm adapted}(x)-f_0(x)$ and measure
$R(x)=\operatorname{RMS}(d(x))$.
\item Return $f_0(x)+\min\{1,\tau/R(x)\}d(x)$ when $R(x)>0$.
If $R(x)=0$, return the source output.
\end{list}
\end{minipage}}
\endgroup
\par\medskip

The cap uses a fixed limit $\tau$ but an image-dependent multiplier.
At $c=.5$, Half and Cap give identical calibration outputs, losses, and
NCG, but can differ on evaluation images. Calibration scores therefore
depend on the output rule. If the calibration correction is zero,
$\tau=0$ and the cap returns the source.

Both rules require the source and adapted outputs and no evaluation
reference. The cap guarantees a source-relative RMSE-change bound in
normalized intensity units, but does not guarantee reconstruction
improvement or a finite PSNR-degradation bound.

\section{Experimental protocol}
\label{sec:protocol}

\subsection{Sources, calibration and evaluation}

We study a fixed fastMRI-to-M4Raw shift: source models are trained on
fastMRI, calibration and evaluation on M4Raw. Seven U-Nets~\cite{unet} and seven ViTs have 70\% sparsity in eligible
tensors. Each architecture has one PUN-IT, one RigL, and five single-image
SNIP (image-SNIP) sources, each using a different pruning slice.
PUN-IT~\cite{punit} learns independent mask-retention probabilities with
initial weights frozen. It minimizes expected reconstruction loss plus an
average Kullback-Leibler divergence (KL) penalty toward the target density, using Gumbel-softmax
approximations, then keeps the largest probabilities.
RigL~\cite{rigl} periodically replaces small-magnitude active weights with
high-gradient inactive connections during training.
For source reconstruction loss $\mathcal L_{\rm src}$,
image-SNIP~\cite{snip} keeps the largest $|w_i\partial\mathcal L_{\rm src}/\partial w_i|$
scores from one source-training slice at initialization.
PUN-IT and image-SNIP fix their masks during subsequent weight training.

Source training uses fastMRI brain training and validation
directories~\cite{fastmri,fastmri_brain}, Adam, L1 loss, batch size one,
and seed 42. U-Net trains for 50 epochs at initial rate $5\times10^{-4}$
and ViT for 40 epochs at $10^{-4}$. Source-validation PSNR selects
checkpoints. U-Net has four pooling stages, 32 base channels, and 21
eligible convolutions. ViT has ten blocks, 16 heads, width 704,
$10\times10$ patches, and 20 eligible MLP maps. Its wrapper standardizes
inputs and restores their mean and standard deviation.

In both domains, coils are inverse transformed, center-cropped to
$320\times320$, and combined by root sum of squares (RSS), separately
for real and imaginary components. A centered fast Fourier transform
(FFT) precedes one shared Cartesian mask with 80 of 320 lines.
Inverse transformation, magnitude, and min-max normalization produce the
inputs. Source RSS targets are normalized before center cropping.

All sources use the same 38 M4Raw~\cite{m4raw} calibration episodes from
seven participant groups, disjoint from 22 evaluation groups. Calibration
uses zero-based slice 9. Frozen adapters evaluate 318 slices from 106
volumes at indices 5, 9, and 12, the rounded 25\%, 50\%, and 75\%
positions after excluding two end slices. Both calibration and evaluation include T1, T2, and FLAIR images.

\subsection{Fitting and statistical setup}

Fitting uses labeled-pair L1 and 20 Adam steps at $10^{-3}$, with
$\beta_1=.9$, $\beta_2=.999$, $\epsilon=10^{-8}$, no weight decay, and
no running-maximum second moment. We retain the earliest minimum-loss
step, with $10^{-12}$ tie tolerance. Step 0 is the source for
zero-initialized adapters. Normalization and dropout remain in evaluation
mode. LoRA initializes $A$ uniformly within $\pm1/\sqrt{d^{\rm in}}$ and
$B=0$, where $d^{\rm in}$ is the input dimension. Three adaptation seeds
reuse the same sources and participants. Support and step selection use
calibration data only.

The association analysis correlates NCG with each adapter's mean evaluation
PSNR gain. Pearson and Spearman coefficients are computed across episodes
within each architecture, pruning family, adapter, and budget, then averaged
over the three pruning families with their signs preserved. Reconstruction
means weight all seven sources equally. PSNR uses the target range. Paired
gains average seeds, slices within volumes, volumes within evaluation groups,
and episodes within calibration groups, with equal group weights.

\textbf{Uncertainty.} We bootstrap participant groups with replacement,
keeping each group's slices together, using 50,000 paired resamples and
shared draws across compared methods. Output-control analyses resample the
seven calibration and 22 evaluation groups independently, with source models
and fitting seeds fixed.

Simultaneous intervals use bootstrap maximum-deviation adjustments
within prespecified comparison families. DA-versus-source and
DA-versus-random use absolute centered deviations, while
DA-versus-$S$, DA-versus-LoRA, and output-control comparisons use
standardized maximum deviations (max-$t$). DA versus LoRA adjusts
over both architectures and all budgets.
\textbf{Correlation comparisons} adjust over 24 tests: Half and Cap versus
Full, for Pearson and Spearman, across six architecture-adapter settings.
\textbf{Quality comparisons} adjust over 36 tests: all three pairs among
Full, Half, and Cap, for mean PSNR gain and the fraction of adapters with
negative mean gain, across the same six settings. These groups are adjusted
separately. The intervals describe stability on this panel, with uncertainty
limited by the seven calibration groups.

\section{Results}

\begin{table*}[t]
\centering
\caption{\textbf{Reconstruction changes at matched parameter counts.}
Mean PSNR change in dB $\pm$ simultaneous 95\% interval half-width.
Bold marks intervals excluding zero before rounding.
$D$ is DA-SNIP and $S$ is unrestricted sparse adaptation.
Comparison groups are adjusted separately (Section~\ref{sec:protocol}).}
\label{tab:results}
\setlength{\tabcolsep}{5pt}\renewcommand{\arraystretch}{1.00}
\begin{tabular}{@{}lcccccc@{}}
\toprule
& \multicolumn{3}{c}{U-Net} & \multicolumn{3}{c}{ViT} \\
Contrast & 0.1\% & 1\% & 10\% & 0.1\% & 1\% & 10\% \\
\midrule
$D-$source & $+.430\pm.442$ & $\mathbf{+.655\pm.442}$ & $\mathbf{+.468\pm.442}$ & $\mathbf{+.509\pm.442}$ & $\mathbf{+.675\pm.442}$ & $\mathbf{+.826\pm.442}$ \\
$D-$random DA & $\mathbf{+.396\pm.327}$ & $\mathbf{+.364\pm.327}$ & $-.127\pm.327$ & $\mathbf{+.488\pm.327}$ & $\mathbf{+.465\pm.327}$ & $+.134\pm.327$ \\
$D-$LoRA & $-.137\pm.216$ & $+.119\pm.146$ & $-.015\pm.240$ & $\mathbf{+.329\pm.180}$ & $+.030\pm.048$ & $0.000\pm.056$ \\
$D-S$ & $\mathbf{-.332\pm.263}$ & $-.156\pm.193$ & $+.031\pm.464$ & $\mathbf{-.200\pm.119}$ & $\mathbf{-.125\pm.079}$ & $+.065\pm.216$ \\
\bottomrule
\end{tabular}

\end{table*}
\begin{table*}[t]
\centering
\caption{\textbf{Output control at 10\% adaptation budget.}
Full, Half and Cap reuse the same fitted adapters. PSNR gains are in dB.
Correlations relate each rule's own NCG to mean evaluation gain.
\textbf{Bold correlations} have a 95\% adjusted interval above zero for
their increase over Full (24 comparisons). Negative adapters have mean gain below zero.
Quality intervals are adjusted separately across 36 comparisons.}
\label{tab:cap-control}
\setlength{\tabcolsep}{4.5pt}
\begin{tabular}{@{}lrrrcccc@{}}
\toprule
& \multicolumn{3}{c}{PSNR gain} & \multicolumn{3}{c}{Pearson / Spearman} & Negative adapters (\%) \\
Model & Full & Half & Cap & Full & Half & Cap & Full / Half / Cap \\
\midrule
U-Net $D$ & 0.468 & 0.684 & 0.745 & -0.380/-0.234 & \textbf{0.526}/\textbf{0.648} & \textbf{0.485}/\textbf{0.556} & 18.60/2.81/3.36 \\
U-Net $S$ & 0.437 & 0.837 & 0.892 & -0.593/-0.417 & \textbf{0.338}/\textbf{0.464} & \textbf{0.472}/\textbf{0.547} & 19.06/1.64/1.35 \\
U-Net LoRA & 0.483 & 0.594 & 0.630 & -0.259/-0.109 & 0.602/\textbf{0.655} & \textbf{0.570}/\textbf{0.627} & 12.91/1.74/2.37 \\
ViT $D$ & 0.826 & 0.582 & 0.761 & 0.412/0.503 & 0.842/0.851 & 0.861/0.871 & 0.09/0.00/0.00 \\
ViT $S$ & 0.761 & 0.692 & 0.753 & -0.044/0.051 & \textbf{0.701}/\textbf{0.717} & \textbf{0.704}/\textbf{0.723} & 2.44/0.00/0.00 \\
ViT LoRA & 0.826 & 0.569 & 0.755 & 0.615/0.698 & 0.858/0.828 & 0.872/0.886 & 0.00/0.00/0.00 \\
\bottomrule
\end{tabular}
\end{table*}

\subsection{Dormant adaptation}

Small dormant supports can improve reconstruction, showing that pruned
weights retain useful adaptation capacity even when only a small fraction
is reactivated. At the 0.1\% budget, DA-SNIP uses 0.043\% of the U-Net
dormant pool and 0.038\% of the ViT dormant pool, with mean gains of
$.430$ and $.509$ dB, respectively. The U-Net interval includes zero.
U-Net gain peaks at 1\%, while ViT gain increases across budgets.

Support selection matters most at small budgets. All four DA-SNIP versus
random DA comparisons at 0.1\% and 1\% favor DA-SNIP
(Table~\ref{tab:results}). All four corresponding point estimates favor
unrestricted $S$, with adjusted support in three comparisons. Dormant restriction therefore provides
a useful compact adaptation space, but does not generally outperform
optimization over all eligible coordinates.

DA-SNIP exceeds LoRA by $.329$ dB $[.149,.510]$ on ViT at 0.1\%, while
the other five DA-SNIP versus LoRA intervals include zero. At this ViT
budget, LoRA updates only three of twenty MLP maps. These comparisons use
one fixed learning rate and rank allocation and do not establish
best-tuned LoRA performance.
The dormant restriction also changes calibration-transfer behavior.
Averaged over the six architecture-budget settings under Full,
Pearson/Spearman correlations are $.33/.41$ for DA, $.01/.07$ for
$S$, and $.33/.34$ for LoRA. At 10\% on U-Net, DA-SNIP also has
higher median alignment than $S$ ($t=.86$ versus $.79$;
Table~\ref{tab:hneg}), suggesting that the restricted update space can
make calibration fit more informative about transfer.
\subsection{Output control}
\label{sec:response}
\label{sec:output-scale}

\begin{figure}[t]
\centering
\includegraphics[width=\columnwidth]{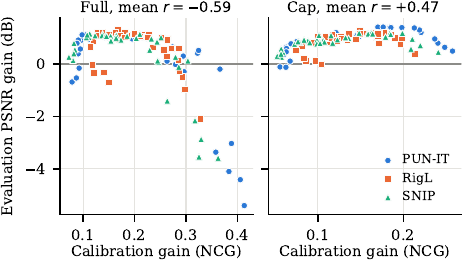}
\caption{\textbf{Calibration fit versus transfer for U-Net $S$ at 10\% budget.}
Each point is one calibration episode, averaged over a pruning family's
sources and three adaptation seeds. Left: Full output. Right: Cap
($c=.5$), with NCG measured on the capped calibration output. Titles
report mean Pearson correlation over pruning families. Table~\ref{tab:cap-control}
reports all six architecture-adapter settings.}
\label{fig:ncg}
\end{figure}

Under Full at 10\% on U-Net, better calibration fit is associated with
worse transfer for all three adapters: Pearson is $-.380$ for DA-SNIP,
$-.593$ for $S$, and $-.259$ for LoRA. Output scale changes both this
association and reconstruction quality. For DA-SNIP, Half raises
Pearson to $.526$ and Spearman from $-.234$ to $.648$, while increasing
mean gain by $.216$ dB. On ViT, however, Half lowers DA-SNIP mean gain
by $.244$ dB, so fixed shrinkage does not give the same quality tradeoff
across architectures.
Figure~\ref{fig:ncg} illustrates this behavior for unrestricted sparse
adaptation on U-Net: under Full, better calibration fit is negatively
associated with evaluation gain, whereas Cap restores a positive
association.

\begin{table}[t]
\centering
\caption{\textbf{Image-level harm at 10\% budget.}
Weighted share (\%) with PSNR below the source ($H(a)<0$), and median
alignment ratio $t$. Full harms for $t<1/2$ and Half for $t<1/4$.}
\label{tab:hneg}
\setlength{\tabcolsep}{3pt}
\renewcommand{\arraystretch}{1.0}
\begin{tabular}{@{}lcccc@{}}
\toprule
Model & Full & Half & Cap & median $t$\\
\midrule
U-Net $D$    & 29.2 & 15.7 & 17.3 & .86\\
U-Net $S$    & 29.3 & 13.9 & 14.1 & .79\\
U-Net LoRA   & 27.5 & 14.5 & 16.6 & .87\\
ViT $D$      & 16.2 &  9.4 & 12.8 & 1.30\\
ViT $S$      & 22.9 & 11.3 & 14.5 & .95\\
ViT LoRA     & 14.3 &  7.6 & 11.2 & 1.23\\
\bottomrule
\end{tabular}
\end{table}

The image-level results connect the scaling analysis to observed failures.
Full harms 27.5--29.3\% of U-Net images, compared with 14.3--22.9\% on
ViT. Median alignment is also higher on ViT ($t=.95$--$1.30$ versus
$.79$--$.87$ on U-Net), consistent with larger useful output scales.
Half reduces harmful-image rates in all six settings, while Cap lies
between Half and Full on this criterion.

The calibration-based Cap addresses the limitation of a fixed scale by
adapting its multiplier to each image's correction size. For DA-SNIP it applies on average 63\% of the fitted correction on
U-Net and 81\% on ViT. Across DA-SNIP,
$S$, and LoRA, both Half and Cap increase Pearson and Spearman relative
to Full in all six architecture-adapter settings
(Table~\ref{tab:cap-control}). Fifteen of the 24 increases have adjusted
intervals above zero. Thus output control consistently strengthens the
measured calibration-evaluation association, although these associations
require validation before use as predictors for new participants.

Cap has higher mean PSNR than Half in all six settings, with advantages
from $.036$ to $.186$ dB and adjusted support in three comparisons.
However, Cap has more harmful image-level outcomes than Half in all six
settings (Table~\ref{tab:hneg}). Thus average quality and image-level
harm need not rank the rules identically. Table~\ref{tab:cap-control}
instead counts adapters with negative mean evaluation gain. Mean PSNR
also depends on source MSE and correction energy through
Eq.~\ref{eq:normalized-gain-psnr}.

\section{Conclusion}
Small dormant supports provide useful single-slice specialization.
We derive how correction alignment and scale determine MSE and PSNR, and
use this analysis to motivate reference-free output control. Across DA,
unrestricted sparse adaptation, and LoRA, output control strengthens the
measured calibration-evaluation association, while the image-dependent cap
enforces a source-relative RMSE-change bound and improves mean PSNR over
fixed half scaling. Main limitations are that the evidence is constrained to fixed source checkpoints and one
constructed fastMRI-to-M4Raw shift, with only seven calibration groups.
Independent training runs and additional acquisition shifts remain for
future validation.

\clearpage
\balance

\section{Compliance with Ethical Standards}
This retrospective analysis used released anonymized fastMRI and M4Raw
data, with no new participants or acquisitions.
fastMRI curation was approved by the New York University School of Medicine
Institutional Review Board~\cite{fastmri}.
M4Raw acquisition was approved by Shenzhen Technology University's
Institutional Review Board (SZTULL012021005), with written consent for
anonymized public release~\cite{m4raw}.
\section{Acknowledgments}
No funding was received for conducting this study. The authors have no
relevant financial or nonfinancial interests to disclose.
\bibliographystyle{official_template_2027/IEEEbib}
\bibliography{references}
\end{document}